\documentclass{aa}  

\usepackage{graphicx}
\usepackage{txfonts}
\begin{document} 

   \title{Wavelet theory applied to the study of spectra of trans-Neptunian objects}
   %\subtitle{I. Overviewing the $\kappa$-mechanism}

   \author{Souza-Feliciano, A.C.\inst{1},
          Alvarez-Candal, A.\inst{1},
          \and
          Jim\'enez-Teja, Y.\inst{1}
          }

   \institute{Observat\'orio Nacional, street General Jos\'e Cristino 77, 20921-400, Rio de Janeiro, Brazil\\
              %T\"urkenschanzstrasse 17, A-1180 Vienna\\
              \email{carolinaastro@on.br}
         %\and
         %    University of Alexandria, Department of Geography, ...\\
         %    \email{c.ptolemy@hipparch.uheaven.space}
          %   \thanks{The university of heaven temporarily does not
           %          accept e-mails}
             }

   %\date{Received September 15, 1996; accepted March 16, 1997}

% \abstract{}{}{}{}{} 
% 5 {} token are mandatory
 
  \abstract
   {Reflection spectroscopy in the near-infrared (NIR) is used to investigate the surface composition of trans-Neptunian objects (TNOs). 
   In general, these spectra are difficult to interpret due to the low apparent brightness of the TNOs, causing low signal-to-noise ratio even in spectra obtained with the largest telescopes available on  Earth, making it necessary to use filtering techniques to analyze and interpret them.}
   {The purpose of this paper is to present a methodology to analyze the spectra of TNOs. Specifically, our aim was to filter these spectra in the best possible way: maximizing noise removal, while minimizing the loss of signal.}
   {We used wavelets to filter the spectra. Wavelets are a mathematical tool that decompose the signal into its constituent parts, allowing us to analyze the data in different areas of frequencies with the resolution of each component tied to its scale. To check the reliability of our method, we compared the filtered spectra with the spectra of water and methanol ices to identify some common structures between them.}
   {Of the 50 TNOs in our sample, we identify traces of  water ices and methanol in the spectra of several of them, some with previous reports, while for other objects there were no previous reports.}
   {We conclude that the wavelet technique is successful in filtering spectra of TNOs.}


\titlerunning{Wavelets applied to TNO spectra}
\authorrunning{Souza-Feliciano et al.}
\maketitle

\section{Introduction}

 Trans-Neptunian objects (TNOs) are icy relics left over from the planetary accretion disk that orbit the Sun beyond Neptune. The investigation of their properties, as remnants of the external planetesimal swarms, is essential to understanding their formation and  evolution. Investigating their surfaces by spectroscopy is a tool for constraining the chemical properties and the evolution processes that occurred in the solar system \citep{barucci2008}. 

To better understand the objects in this region we rely on observations, mostly ground-based, of their surface properties. In particular, we use reflection spectroscopy in the near-infrared (NIR). In general, the NIR spectra of TNOs are difficult to interpret because of the objects' low apparent brightness, which  causes a low signal-to-noise ratio \textbf{(S/N)} even in data obtained with the largest telescopes available on the Earth. Therefore, the spectra of these objects usually have a considerable amount of noise making it necessary to use filtering techniques to analyze and interpret them. 

In signal processing, noise is defined as that undesirable, random signal that is mainly introduced by the processes of capture, storage, and transmission of the data. Although inherent to the data and thus difficult to be disentangled from them, the noise must be estimated and removed in the most efficient way without altering the real information contained in the data. Hereafter we  use the term signal to indicate that part of the data containing real information, and noise to describe that artificial, random signal introduced by the mechanisms previously described.

Filters are usually applied to remove noise from signals. Due to its random nature, noise is identified with unstructured high frequencies. Filtering techniques try to identify these frequencies and eliminate or attenuate them.

Among the most commonly used techniques are rebinning, the runing box, and the Fourier transform. However, these techniques are not completely efficient because they either decrease the spectral resolution (rebinning and runing box) or they can incorrectly remove high frequency spectral features easily confused with noise (Fourier transform). Two important problems affect this last technique: 1) the definition of ``high frequency'' is somewhat arbitrary, and lacks a robust definition, and 2) data are processed as a whole. In this sense, the Fourier transform provides the frequencies that globally compose the data. There is no information on the localization of these frequencies, that is, whether they appear throughout the whole signal or just in a certain interval. As they are all treated equally with this global analysis, it is difficult to determine whether a high frequency is truly related to noise or instead a real feature of the signal, thus containing information. This second problem is usually  overcome by applying the Fourier transform in windows, but the arbitrary definition of the size of the window and the border effects induced by them made us choose a different technique to filter our TNO data.

Wavelets are similar to the Fourier transform since they decompose the data into the frequencies that compose them, but with the difference that these are localized (although not perfectly). The wavelets perform a multiscale analysis of the data, where the different frequencies that compose the signal are roughly located with an uncertainty that depends on the resolution of the scale. The information provided by the wavelets is thus essentially the same as that from Fourier, but detailed and split between different scales. With this new approach, a high frequency is not immediately associated with noise, but depending on its modulus (i.e., its presence and weight in the data), it will be considered a significant piece of information or noise.

Wavelets have been used in astrophysics for various purposes: the denoise cosmological simulations \citep{2003romeo, 2004romeo}, to remove the background light in galaxy clusters and detect faint small objects to complete their luminosity function \citep{2017live}, to identify the different scales in wavelet decomposition with luminous components in groups of galaxies \citep{darocha08, mendes}, and to study mutual events among Galilean satellites \citep{ramirez2006new}, among others.

Here we show how the application of the wavelets to filter the data allows us  to reduce the amount of noise in the spectra of 50 TNOs obtained with the instrument SINFONI at the Very Large Telescope in Cerro Paranal, Chile. We  describe the sample in the Sect. 2  and the chosen fitering technique in Sect. \ref{processing}. The methodology used is shown in Sect. \ref{metho} and the results are presented in Sect. \ref{resultados}. Finally, the conclusions are drawn in the Sect. 6.

%__________________________________________________________________

\section{Data}\label{analyses}

%                                     Two column figure (place early!)
%______________________________________________ Gamma_1 (lg rho, lg e)
All spectra shown in this paper were obtained with the Spectrograph for Integral Field Observations in the Near-Infrared  (SINFONI)\footnote{http://www.eso.org/sci/facilities/paranal/instruments/sinfoni.html} located in the Cassegrain focus of the unit 4 of the Very Large Telescope (VLT). SINFONI  provides a spectroscopic imaging in 3-D in the spectral range between 1.05 and 2.45 $\mu$m. This instrument has four different modes of observation, J, H, K, and H+K, with a spectral resolution of 1500 in the H+K band. The spectra of the sample were obtained using the H+K mode because it allows us to observe the H and K bands simultaneously. The \textbf{spatial }resolution was 0.25 arcsec/spaxels, with corresponding field of view of 8.0 arcsec. 

Our sample was selected from the objects observed within an ESO Large Program carried out between 2007 and 2009 (P.I.: M. A. Barucci). The objects analyzed are listed in Table 1. Most of the spectra presented here have already been published. Details of those marked with (+) can be found in \cite{barucci2011new}, with (++) in \cite{demeo2010spectroscopic}, with (--) in \cite{barucci201090377}, with (\#) in \cite{aham}, with (*) in \cite{guilbert2009}, with (-) in \cite{alvarez07}, (**) in \cite{merlin2007}, and Pluto and Charon in \cite{MERLIN2010930}. We note that the spectra of Chiron and 1996 TO$_{66}$ have not been previously published; nevertheless, the details of their reduction can be found in any of the previous references. In this way, according to the classification of \cite{glad}, among the spectra in our sample are included 13 classical objects, 10 resonant objects, 16 centaurs, 4 scattering disk objects, and 5 detached objects.

The sizes of the objects range from 20 to 2700 km (see Table 1),  several of which with diameters over 500 km (Pluto, Triton, Haumea, Quaoar, etc.). From a technical point of view, it is easier to detect absorption features in the spectra of the largest objects, due to their S/N and also because large objects can retain a larger reservoir of volatiles \citep{schaller}. Smaller objects, which constitute most of our sample, are more difficult to deal with because (1) they are too small to have kept the volatiles that were accreted onto them when they were formed \citep{schaller}; and/or (2) of their small sizes, their apparent brightness is too faint and therefore the S/N is too low to detect small absorption features.
 
Unfortunately, nothing can be done in case (1), but we can tackle  case (2) and try to work with the spectra generated by these objects through the application of techniques that improve their signal-to-noise ratios until the arrival of the next generation of telescopes. Therefore, we applied the wavelets to filter all objects in our sample. For large TNOs (d $> $ 500 km) \citep{brown2012}, especially those that have been well studied in the literature, the goal was to verify whether  the application of wavelets can be a tool for reducing the noise and for recognizing the absorption bands already reported in the literature without losing information. As the result was positive, we applied the same technique to smaller TNOs (d $ <$ 500 km) and tried to identify possible evidence of absorption bands of water and methanol ices that the noise could be hiding. 

\section{Data processing}\label{processing}

As mentioned above, due to the overall faintness of the TNOs, their spectra usually have relatively low S/N (seldom over 20 or 30). It is therefore necessary to use a filtering technique to analyze and interpret them. 

The main purpose of our study is to use the wavelets technique to reduce the maximum amount of noise without changing the shapes and positions of the possible absorption bands of the spectra, if present, in order to identify them. Once filtered, some spectral parameters are calculated to estimate the possibility of absorption detection, due to some ices of astrophysical interest such as water and methanol ices. After that, these spectra are compared to spectra calculated from optical constants measured in the laboratory for the ices mentioned. This step is necessary to link the absorptions detected in the filtered spectra with the ices we are looking for.  

\subsection{Wavelets}\label{wav31}

The wavelet transform is a multiresolution technique which decomposes the data into their constituent frequencies, roughly located in time (or wavelength) and frequency simultaneously. It is based on the so-called wavelet function or mother wavelet $\psi$, which can be any squared-integrable function with zero mean:

\begin{equation}
 \int_{-\infty}^{+\infty}\psi(t)dt = 0.
\end{equation}

These two simple properties guarantee that $\psi$ has an undulating shape, and is the reason why it is called a wavelet. For each wavelet $\psi$, we can build its associated wavelet family, consisting of the original wavelet function $\psi$ dilated and translated by different factors, the scale $j$ and the translation factor $k$, respectively:

\begin{equation}
\psi_{j,k}= \frac{1}{\sqrt{2^j}}\psi\left(\frac{t-2^jk}{2^j}\right).
\end{equation}

It can be proved that these wavelet families are mathematically orthonormal bases of the $\mathcal\{L\}^2$-space of the squared integral functions. This means that any function (or data), provided that it is smooth enough, can be decomposed as the linear combination of the elements of the wavelet family

\begin{equation}\label{decomp}
f(t)= \sum_{j=0, j \in \mathbb{Z}}^{\infty} \sum_{k \in \mathbb{Z}} d_{j,k}\psi_{j,k}(t),
\end{equation}
where the coefficients $d_{j,k}$ are the wavelet coefficients. These coefficients are calculated according to the following expression:

\begin{equation}
d_{j,k}= \int_{-\infty}^{+\infty} f(t) \overline{\psi_{j,k}(t)}dt.
\end{equation}

The  wavelet coefficients are related to the details (frequencies) that compose the data. Their lower scales ($j$) give information on the finer details that compose the data, such as high frequency structures, pseudo-impulsive features, or noise. These fine details are very well localized in time due to the small support size of the associated wavelet functions $\psi_{j,k}$ (its width in the time space). Higher  $j$ scales tell us about coarser details, low frequency structure in the data, usually related with information. Due to the increase in the support size of the $\psi_{j,k}$, these coarse details are localized in time with a higher uncertainty. However, the support size of a function in time is inversely proportional to its support size in the Fourier domain. That means that the narrower the $\psi_{j,k}$ are, the wider their support in frequency becomes, thus increasing the uncertainty in frequency. For this reason, in the wavelet decomposition the high frequencies that compose the data are not very well identified, although exceptionally well localized in the data (i.e., we know where they are, but not who they are), whereas the low frequency structures are better defined, but we do not know exactly where they appear (i.e., we know who they are, but not where they are). So, if Fourier transform provides the exact decomposition in frequency of the data independently of its localization, the wavelet transform tells  us approximately which frequencies compose the data and approximately the intervals where they appear. We lose precision in frequency to gain some information on time.

This is very useful for  disentangling the noise from the signal. In the wavelet theory, the noise is no longer related to the high frequencies, but to those structures that do not have a strong weight in the decomposition in a certain time interval. For example, the high frequency components of a narrow absorption line would be kept as signal (information) since these high frequencies contain structural information of the data in that interval. However, these same high frequencies can be identified as noise in any other interval of the data where the subjacent structure is smoother, like a broad line in a noisy spectrum, since the weight of these high frequencies is small compared to that of the low frequencies. Thus, the noise will be defined as the high frequency in a certain interval which does not have a relative importance for the structure of the data in that interval, identified by the modulus of its associated wavelet coefficient $d_{j,k}$. To quantify this idea, \cite{tres} proved that when there is Gaussian noise, there is a certain threshold (called the universal threshold) below which a wavelet coefficient has a  very high statistical probability of carrying noise. This universal threshold for a signal of size $N$ polluted with white noise of variance $\sigma^2$ is defined as

\begin{equation}
T = \sigma \sqrt{2 \log N }.
\end{equation}

The only problem of the universal threshold is that it assumes the variance of the noise to be known. As this is not usually the case with real data, it can be proved that the following parameter is a robust estimator of the variance $\sigma^2$ of the noise,

\begin{equation}
\sigma = \frac{M}{0.6745}
,\end{equation}

\noindent{where $M$ is the median of the absolute moduli of the wavelet coefficients in the first scale.}

Thus, to filter our noisy TNO spectra we  decompose them up to a certain scale $j$ using a certain wavelet family, calculate the corresponding universal threshold $T$, and set to  zero all the coefficients with $|d_{j,k}|<T$. This is called a ``hard filter'' and it guarantees that the characteristics of the signal, in our case the shape and depth of the absorption bands in our spectra, are preserved after removing the noise.

\subsection{Selection of the wavelet}

There are several wavelet functions (each  with its associated wavelet family) that lead to different wavelet decompositions. The wavelets functions can be real or imaginary, have peaks or be smooth, be symmetric or asymmetric, be more or less oscillating, etc. Therefore, there is no ``perfect'' wavelet function; we must choose the one that works best for our problem according to two main criteria, the characteristics of the data and our final purpose.

As the wavelet decomposition is just the result of the convolution of the data with the mother wavelet, dilated and translated by different factors, wavelet functions that are similar to the data perform better. In this sense, if our data have symmetric features (e.g., the absorption lines in our spectra) a symmetric wavelet is  preferred. A smooth, not very oscillating wavelet would also be desirable in our case, since the spectra do not present sharp peaks or highly oscillating features (e.g., a seismic signal).

The final goal of the wavelet decomposition is also very important for the selection of the wavelet. Wavelets can be used to detect singularities in the data, for instance sharp peaks, extrema, inflection points (see, e.g., \citealt{ramirez2006new}). For this purpose, the support size of the wavelet is crucial since wavelets with wide support introduce too much uncertainty in time, and the singularity will not be tightly identified. When we are interested in filtering the data, the support size of the wavelet is not essential (although a narrow wavelet function is always desirable), but  its compression capacity is much more important. The wavelet transform keeps the energy of the data, meaning that the norm of the signal is the same as the norm of the wavelet coefficients of  its decomposition. As the filtering algorithm is based on the comparison of the modulus of the wavelet coefficients carrying noise with the modulus of the coefficients associated with the signal, a large difference between these two sets would be convenient. If the energy of the signal were split into many coefficients, this comparison might not be clear and a coefficient containing signal could be misidentified as noise. However, if the energy of the information were concentrated in just a few wavelet coefficients, the filtering process would be straightforward. The concept of vanishing moments helps us  know the capacity of concentration (compactness of the decomposition) that has the wavelet. In practice, the longer the number of vanishing moments that the function has, the fewer the coefficients needed in the decomposition (Eq. \ref{decomp}).

Therefore, to select the wavelet that best filters our data, up to a certain scale $j$, we can follow two paths: (i) perform the filtering for different wavelets and different scales for each input spectrum one by one and decided manually which is the best or (ii) devise an automatic algorithm. The first case might work for one or two spectra, but it is highly inefficient for a larger number, therefore we will follow path (ii).

 We devised a blind test in order to scan among the discrete wavelets offered by the PyWavelet package of Python with changing scales between 1 and 8. As a test sample, we used  41 spectra of water ice with signal-to-noise ratios ranging from about 1 to about 200, with emphasis in the range 1--50.

In order to select the optimal wavelet we use a quadratic sum of three different figures of merit (FoMs, taken from \citealt{meza2010quantitative}). All of them are blind, i.e., with no a priori information regarding the data (e.g., what ice lies underneath the noise.)

Before continuing we  define our input spectrum (data) as $I=[I_i,...,I_M]$ and the filtered spectrum (signal) as $F=[F_i,...,F_M]$.

The first FoM is the roughness (adapted from \citealt{hayat1999statistical}), 
\begin{equation}
        {\rm \rho} = \sum_{i=1}^{M}{|[-1,1]*F_i|}/\sum_{i=1}^{M}{|F_i|}
,\end{equation}
which measures the roughness of a signal. The asterisk stands for convolution. A constant, smooth signal will have $\rho=0$, while increasing pixel-to-pixel variation will have larger values of $\rho$.

A second FoM is the noise reduction (NR, adapted from \citealt{chen2003destriping}),
\begin{equation}
        {\rm NR = N_0/N_1},
\end{equation}
where N$_0$ is the average of the power spectrum of the data and N$_1$ is the average of the power spectrum of the signal. The best filtering is obtained when NR is high.

The last FoM we use is the residual non-uniformity (RNU, adapted from \citealt{wang2008enhanced}),

\begin{equation}
        {\rm RNU} = {\rm 100/\overline{F}\sqrt{(1/M\times \sum_{i=1}^{M}{(F_i-\overline{F})^2})}},
\end{equation}
where we search for the most uniform filtered signal, and  therefore for  low values of RNU. The overlined quantities are mean values.

In order to define our metric we first invert NR to search for its minimum and renormalize all FoMs to the interval [0,1].
Then we define our blind figure of merit (bFoM) as

\begin{equation}
       {\rm bFoM} = \sqrt{\sum_{i=1}^{3}({\rm FoM_i})^2/3} ;
\end{equation}
as such, the optimal wavelet and scale for filtering will be given by the minimum of bFoM.

\begin{figure*}[!ht]
\centering
\includegraphics[height = 8cm]{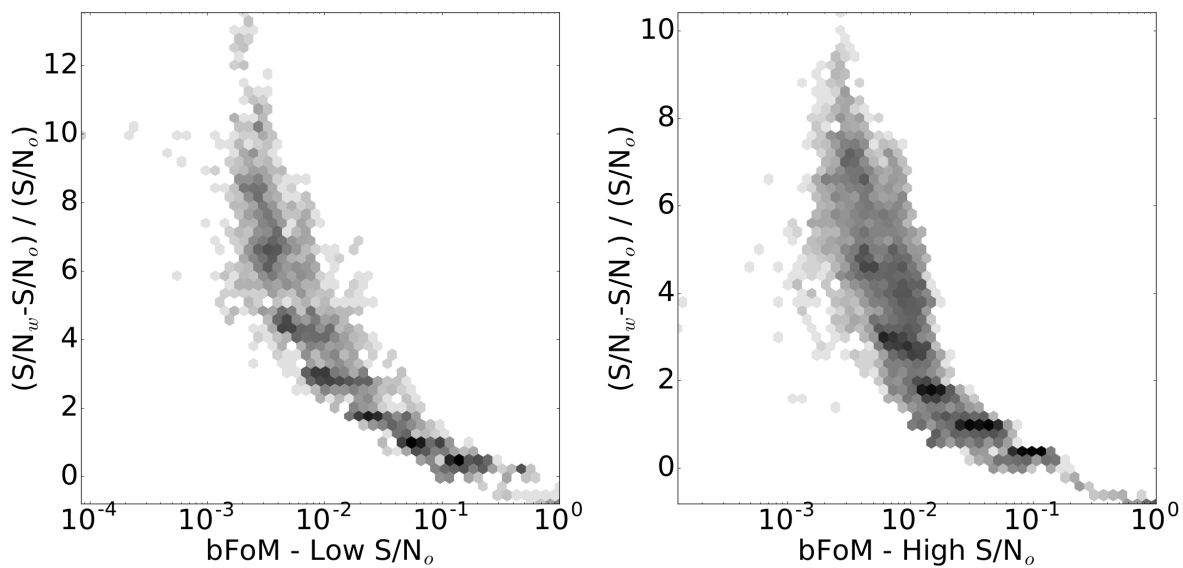}
\caption{Maps of the distribution of the bFoM compared to the relative gain in S/N$_w$. The right panel shows data with S/N$_o$ < 20, while the left panel shows data with S/N$_o$ > 20. (See text.) }
\label{bfom}
\end{figure*}

By construction we know the input signal-to-noise ratio, S/N$_o$, of the spectra in the test sample. Each one was filtered with 58 different wavelets, in eight different scales, amounting to 454 filtered spectra per input spectrum. Considering that we have 41 modeled spectra, in the end we have 19014 filtered spectra each with a value of bFoM. To help visualize the results we created maps that show a clear trend to larger gains in S/N with decreasing values of bFoM (Fig. \ref{bfom}). Therefore, we adopt as our criterion for selecting the wavelet and scale the one that minimizes bFoM; as such, we are selecting the wavelet and scale that optimally balances  the smoothness, uniformity, and noise removal of the data. Upon visual inspection of the signal (the filtered spectrum), the results are satisfactory.

It must be mentioned that although low values of bFoM ensure good gains in S/N, it does not necessarily maximizes it; this is a trade off between knowing a priori what the ``noiseless'' spectrum should look like and no a priori assumptions.

Figure \ref{bfom} has been split into two panels. The left panel shows the maps for model spectra with S/N$_o$<20, while the right panel does it for S/N$_o$>20. We have chosen this
value as the average S/N of the spectra in our sample (see Table 1) and to illustrate that the performance of bFoM is not correlated by the S/N$_o$ of the spectrum.

\begin{figure}[!ht]
\centering
\includegraphics[height = 7cm]{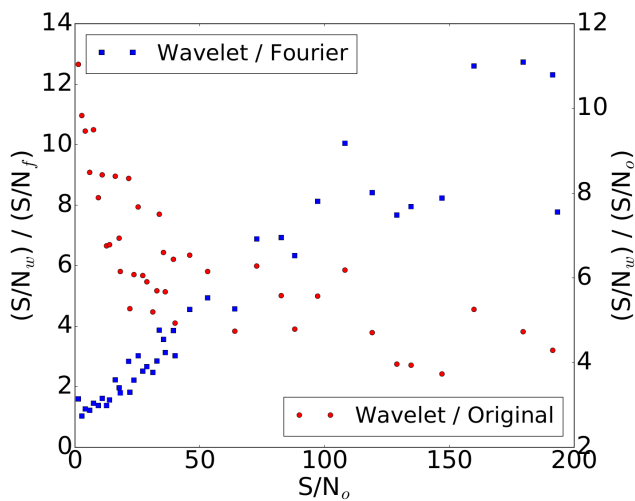}
\caption{Comparison between the performances of the optimal wavelet, found by our algorithm, and a Fourier filtering. The blue squares show the ratio of S/N between wavelet and Fourier filtering, while the red circles show the S/N ratio between the wavelet filtering and the original data.}
\label{S/Ns}
\end{figure}

We also computed the performance of a typical Fourier filtering, applying a fast Fourier transform (S/N$_f$), against the filtering with the optimal wavelet (S/N$_w$). It can be seen in Fig. \ref{S/Ns} that the optimal wavelet clearly outperforms the Fourier filtering even for low S/N$_o$ spectra. It is also clear that S/N$_w$ is at least a factor of 2 better than the S/N$_o$.

Therefore, in what follows, we always use the wavelet and scale that minimizes the bFoM and perform a visual inspection to corroborate the quality of the filtering.

\section{Methodology}\label{metho}

We applied the algorithm mentioned above to our 50 spectra. The signal-to-noise ratio before and after the filtering are shown in Table 1.

In the filtered spectra we measured two absorption band depths, the first at 2.0 $\mu$m, which gives an idea about the presence of water ice, and the second at 2.27 $\mu$m related to methanol (e.g., \citealt{merlin2012}). Fluxes were computed as the median value between 1.98--2.02, 2.18--2.22, and 2.25--2.29 $\mu$m. The error assigned to each value comes directly from the standard deviation of the points used to compute the depth of the absorption band in each interval. These results are presented in Table 1.

Finally, we compare the filtered spectra with laboratory ices of water and methanol
ices, obtained using the \cite{hapke1981} model and optical constants from \cite{quirico1997near}. The objective was to confirm the presence of these ices in the objects where their presence was reported, in order to validate our technique. We also looked for evidence of  these ices in the spectra of TNOs that were previously neglected because of their low signal-to-noise ratios.

\begin{table*}

\centering
 \begin{tabular}{ccccccccc}
 \hline
Number & Wavelet & Scal & D$_{2.0} \mu m$ (\%)  & D$_{2.27} \mu m$ (\%) & Diameter (km)\footnotemark[1] & S/N$_{\rm o}$ & S/N$_{\rm w}$\\
\hline
2060 Chiron & Bior 5.5 & 7 & 3.74 $\pm$ 0.07 & 13.3 $\pm$ 1.3 & 215.6 & 56.1 & 357.6 \\
5145 Pholus + & Rbio 3.9 & 5 & 7.7 $\pm$ 0.4 & 33.9 $\pm$ 9.2 & 119 & 8.5 & 38.2\\
10199 Chariklo * & Bior 5.5 & 5& 6.8 $\pm$ 0.2 & 1.84 $\pm$ 0.09 & 231 & 86.2 & 179.5\\
15874 1996 TL$_{66}$ *,+ & Bior 5.5 & 8 & 22.7 $\pm$ 12.8 & 35.1 $\pm$ 16.6 & 339 & 6.3 & 46.6\\
19308 1996 TO$_{66}$& Bior 5.5 & 7 & 33.6 $\pm$ 40.1 & 33.0 $\pm$ 37.0 & 200 &0.95 & 9.91\\
26375 1999 DE$_{9}$ *& Bior 5.5 & 8 & 9.22 $\pm$ 0.05 & 5.36 $\pm$ 1.04 & 311 & 10.6 & 213.0\\
28978 Ixion * & Rbio 3.9 & 7 & 11.5 $\pm$ 1.5 & 27.9 $\pm$ 16.8 & 617 &16.5 & 92.2\\
32532 Thereus * & Bior 5.5 & 8 & 10.0 $\pm$ 1.0 & 8.9 $\pm$ 1.9 & 62 & 15.7 & 122.2\\
38628 Huya - & Bior 5.5 & 6 & 7.38 $\pm$ 0.43 & 21.0 $\pm$ 2.4 & 458 & 9.0 & 31.6\\
42355 Typhon * & Bior 5.5 & 7 & 14.5 $\pm$ 0.2 & 4.8 $\pm$ 0.8 & 185 & 18.5 & 81.2 \\
44594 1999 OX$_{3}$ + & Rbio 3.5 & 8 & 4.15 $\pm$ 0.85 & 1.46 $\pm$ 0.43 & 135 & 15.7 & 270.0\\
47171 1999 TC$_{36}$ * & Rbio 2.6 & 7 &8.16 $\pm$ 0.74 & 0.74 $\pm$ 0.12 & 393 & 24.7 & 185.0\\
47932 2000 GN$_{171}$ * & Rbio 2.8 & 6 & 20.0 $\pm$ 4.0 & 21.7 $\pm$ 7.40 & 147 & 9.08 & 34.68\\
50000 Quaoar * & Rbio 2.8 & 8 & 28.1 $\pm$ 0.4 & 11.8 $\pm$ 0.5 & 1036 & 23.7 & 28.9\\
52872 Okyrhoe ++ & Sym 15 & 8 & 2.57 $\pm$ 0.07 & 0.72 $\pm$ 0.01 & 35 & 56.7 & 180.6\\
54598 Bienor * & Rbio 2.4 & 8 & 12.7 $\pm$ 1.9 & 25.4 $\pm$ 12.8 & 198 & 6.4 & 96.5\\
55565 2002 AW$_{197}$ * & Db 2 & 8 & 13.8 $\pm$ 7.9 & 21.3 $\pm$ 6.4 & 768 & 16.4 & 190.2\\
55576 Amycus + &Bior 5.5 & 7 & 12.6 $\pm$ 0.4  & 13.5 $\pm$ 2.0  & 104 & 15.2 & 148.8 \\
55637 2002 UX$_{25}$ + & Bior 5.5 & 7 & 12.4 $\pm$ 1.3 & 38.0 $\pm$ 8.8 & 692 & 7.0 & 11.6\\
55638 2002 VE$_{95}$ \# & Bior 5.5 & 8 & 10.28 $\pm$ 0.05 & 11.3 $\pm$ 1.1 & 250 &21.4 & 267.4\\
60558 Echeclus * & Bior 5.5 & 8 & 2.5 $\pm$ 0.4  & 1.2 $\pm$ 0.5 & 65 & 10.1 & 41.8\\
73480 2002 PN$_{34}$ ++ & Bior 5.5 & 8 & 13.9 $\pm$ 2.2 & 2.0 $\pm$ 0.6 & 112 & 12.0 & 95.7\\
83982 Crantor * & Db 18 & 8 & 14.0 $\pm$ 1.8 & 34.3 $\pm$ 11.0 & 59 & 15.8 & 141.9\\
90377 Sedna -& Bior 5.5 & 7 & 25.7 $\pm$ 10.0 & 30.7 $\pm$ 6.2 & 906 & 12.5 & 108.6\\
90482 Orcus * & Rbio 2.6 & 8 & 38.3 $\pm$ 1.3 & 19.6 $\pm$ 1.6 & 958 & 15.9 & 22.5\\
90568 2004 GV$_{9}$ * & Rbio 2.8 & 7 & 9.1 $\pm$ 0.2 & 26.2 $\pm$ 0.5 & 680 & 14.2 & 101.3\\
95626 2002 GZ$_{32}$ + & Rbio 2.4 & 7 & 8.27 $\pm$ 1.25 & 4.87 $\pm$ 1.38 & 237 & 11.5 & 433.8\\
119951 2002 KX$_{14}$ *& Rbio 3.7 & 8 & 7.9 $\pm$ 0.1 & 34.9 $\pm$ 0.7 & 455 & 7.0 & 290.5\\
120061 2003 CO$_{1}$ + & Rbio 2.8 & 7 & 5.16 $\pm$ 0.33 & 0.42 $\pm$ 0.21 & 94 & 29.9 & 52.3\\
120132 2003 FY$_{128}$ * & Sym 13 & 8 & 28.0 $\pm$ 6.0 & 10.8 $\pm$ 13.7 & 460 & 3.0 & 18.0\\
120348 2004 TY$_{364}$ + & Bior 5.5 & 8 & 2.27 $\pm$ 0.02 & 0.60 $\pm$ 0.01 & 512 & 16.0 & 146.8\\
134340 Pluto & Bior 3.1 & 6 & 4.31 $\pm$ 0.12 & 76.0 $\pm$ 17.6 & 2372 & 46.1 & 46.2\\
136108 Haumea ** & Rbio 3.9 & 5 & 59.8 $\pm$ 1.8 & 27.0 $\pm$ 2.6 & 1240 & 13.0 & 14.7\\
136199 Eris * & Bior 5.5 & 6 & 11.6 $\pm$ 0.8 & 92.4 $\pm$ 11.6 & 2326 & 18.46 & 21.2\\
144897 2004 UX$_{10}$ + & Rbio 2.8 & 6 & 18.5 $\pm$ 3.3 & 34.6 $\pm$ 6.9 & 398 & 11.4 & 63.6\\
145451 2005 RM$_{43}$ + & Bior 5.5 & 8 & 21.9 $\pm$ 1.8 & 22.0 $\pm$ 1.9 & 252 & 5.4 & 36.1\\
145452 2005 RN$_{43}$ * & Bior 5.5 & 8 & 1.08 $\pm$ 0.06 & 0.94 $\pm$ 0.01 & 679 & 18.8 & 126.5\\
145453 2005 RR$_{43}$ + & Bior 5.5 & 8 & 76.4 $\pm$ 2.6 & 17.6 $\pm$ 0.3 & 670 & 3.6 & 5.2\\
174567 Varda + & Db 20 & 8 & 7.0 $\pm$ 0.5 & 9.9 $\pm$ 1.5 & 792 & 12.5 & 128.5\\
208996 2003 AZ$_{84}$ +,* & Db 4 & 8 & 20.2 $\pm$ 3.3 & 16.1 $\pm$ 6.7 & 727 & 7.7 & 41.0\\
229762 2007 UK$_{126}$ + & Rbio 2.6 & 8 & 11.9 $\pm$ 1.1 & 10.8 $\pm$ 3.1 & 599 & 10.5 & 95.1\\
250112 2002 KY$_{14}$ + & Bior 5.5 & 8 & 5.0 $\pm$ 0.3 & 7.9 $\pm$ 0.7  & 47 & 19.8 & 110.1\\
281371 2008 FC$_{76}$ + & Bior 5.5 & 8 & 2.88 $\pm$ 0.02 & 13.3 $\pm$ 2.8 & 68 & 10.9 & 231.7\\
307616 2003 QW$_{90}$ * & Bior 5.5 & 8 & 3.85 $\pm$ 0.05 & 12.2 $\pm$ 1.3 & 366 & 2.05 & 11.0\\
309737 2008 SJ$_{236}$ + & Bior 5.5 & 7 & 12.5 $\pm$ 0.3 & 63.0 $\pm$ 52.5 & 18 & 5.6 & 71.8\\
455502 2003 UZ$_{413}$ + & Bior 5.5 & 8 & 25.0 $\pm$ 2.2 & 2.6 $\pm$ 1.8 & 1214 & 7.0 & 130.7\\
2007 UM$_{126}$ + & Rbio 3.7 & 7 & 7.4 $\pm$ 0.2 & 19.4 $\pm$ 6.4 & 42 & 10.4 & 130.0\\
2007 VH$_{305}$ + & Rbio 2.8 & 6 & 13.2 $\pm$ 4.6 & 23.9 $\pm$ 13.8 & 24 & 5.6 & 39.4\\
Charon& Rbio 2.8 & 8 & 53.4 $\pm$ 0.7 & 3.2 $\pm$ 0.1 & 1212 & 9.8 & 9.8\\
Triton\footnotemark[2] & Bior 4.4 & 6 & 16.6 $\pm$ 1.1 & 42.4 $\pm$ 3.0 & 2706 & 51.4 & 54.0\\
 \hline
 \end{tabular}
 \caption{Spectral parameters of observed objects. This table give us an idea about the presence of methanol and/or water ices on the surface of theses objects. We took advantage of the last two columns to show the signal-to-noise ratio before and after the filtering, and in Col. 2 we show which wavelet was used to perform the filtering, where Bior is the wavelet Biorthogonal, Rbio is Reverse biorthogonal, Sym is Symlets, and Db is Daubechies.}

\end{table*}

\footnotetext[1]{The diameters shown here were taken from the survey ``TNOs are cool'' (http://public-tnosarecool.lesia.obspm.fr/).}
\footnotetext[2]{Its believed that Triton had its origin in the trans-Neptunian region, and that it was captured
by Neptune   during the evolution of the solar system.}

\section{Results}\label{resultados}

\subsection{Filtering}

We show here two examples that are representative of our sample: a large and a small body (according to the classification of \citealt{brown2012}). 

\begin{figure}[!ht]
\centering
\includegraphics[height = 4.8cm]{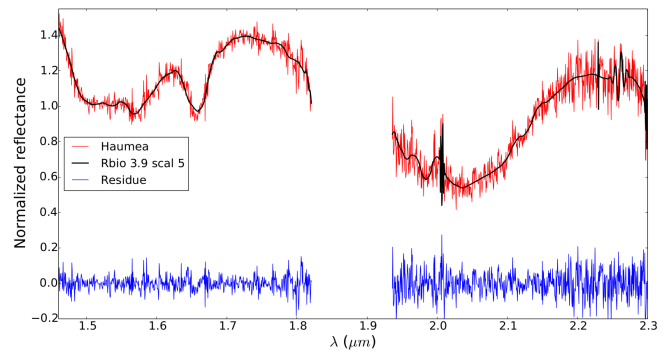}
\caption{Spectrum of 136108 Haumea. The original data is shown in red, the filtered spectrum with the wavelet RBio 3.9 (scale 5) is shown in black, and the difference between the original and filtered spectra is shown in blue.}
\label{il1}
\end{figure}

\begin{figure}[!ht]
\centering
\includegraphics[height = 4.8cm]{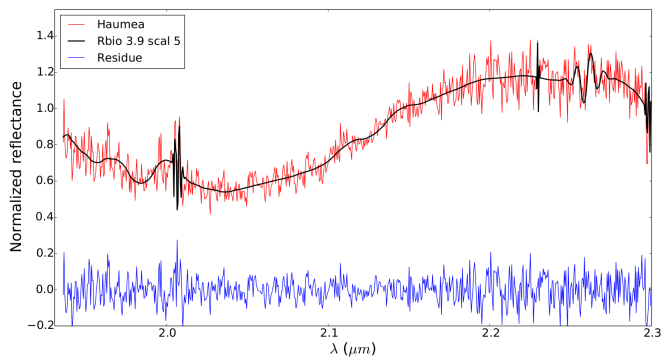}
\caption{A zoom-in of the region between 1.9 and 2.3 $\mu$m of the Haumea spectrum for better visualization of the filtration. The original spectrum (before filtering) is shown in red, the filtered spectrum with the wavelet RBio 3.9 (scale 5) is shown in black, and the difference between the original and filtered spectra is shown  in blue.}
\label{il2}
\end{figure}

Figure \ref{il1} shows the filtering applied to the spectrum of 136108 Haumea, which  is a large TNO with an apparent brightness  high enough to generate a spectrum with a high signal-to-noise ratio. According to the algorithm, the wavelet that best filters this spectrum is the wavelet Reverse Bi-orthogonal 3.9 in the scale 5. The technique was successful in filtering the spectrum because the positions of the absorption bands of the original spectrum were not displaced nor were their shapes altered. Analyzing the residue distribution, we see that there are no structures or patterns, and the signal-to-noise ratio increased. Zooming into the region between 1.9 and 2.3 $\mu$m (Fig. \ref{il2}) where there is  absorption due to water ice, it is possible to perceive with more clarity that the technique did not alter the shape of the band. This indicates that the wavelets only removed noise while respecting the signal of the spectrum.

\begin{figure}
\centering
\includegraphics[height = 4.8cm]{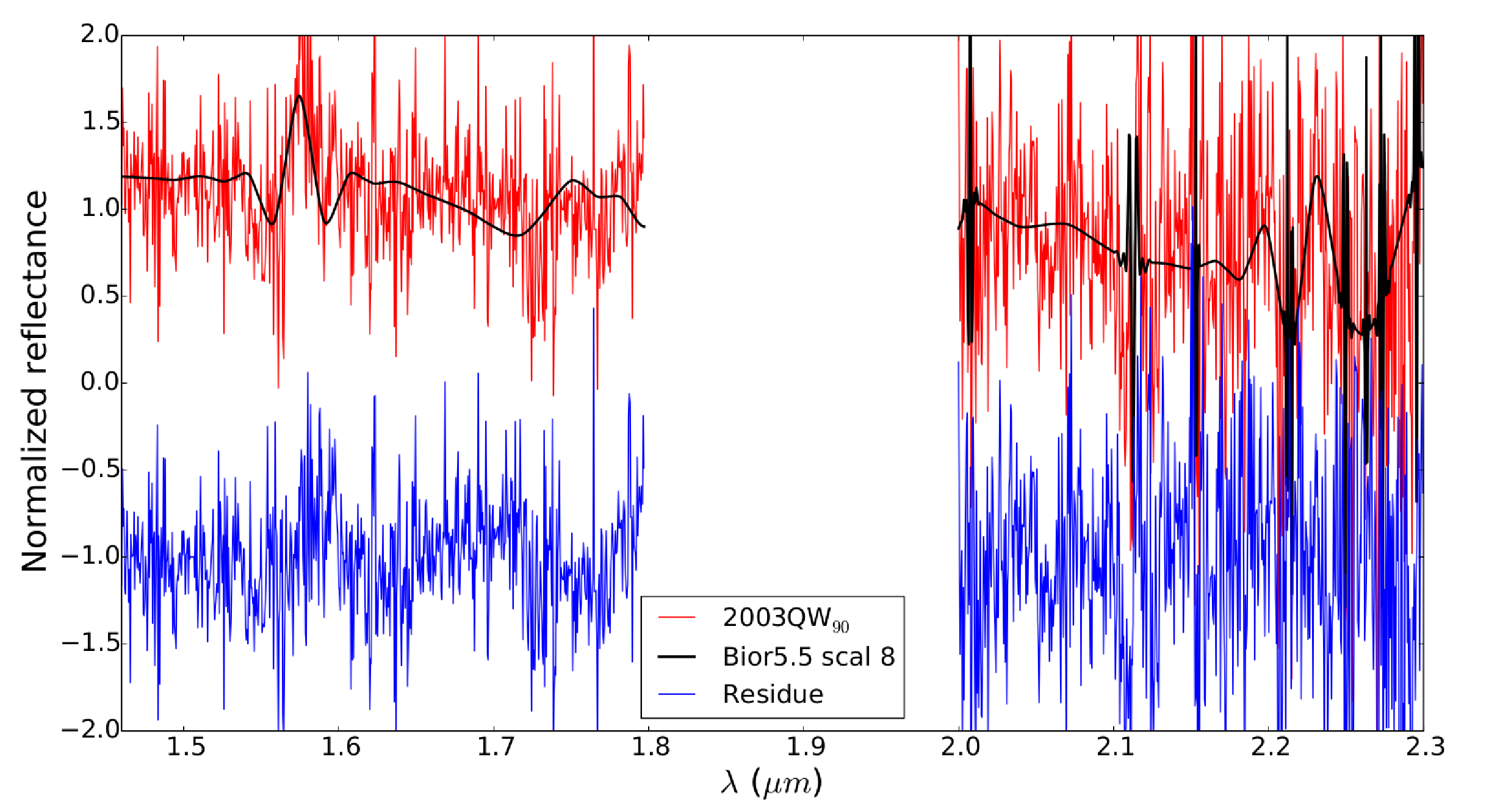}
\caption{Spectrum of 307616 2003 QW$_{90}$. The original data is shown in red, the filtered spectrum with the wavelet Bior 5.5 (scale 8) is shown in black, and the difference between the original and filtered spectra is shown in blue. The residue distribution was shifted downward by 1 unit for better visualization. }
\label{tno_p}
\end{figure}

\begin{figure}
\centering
\includegraphics[height = 4.8cm]{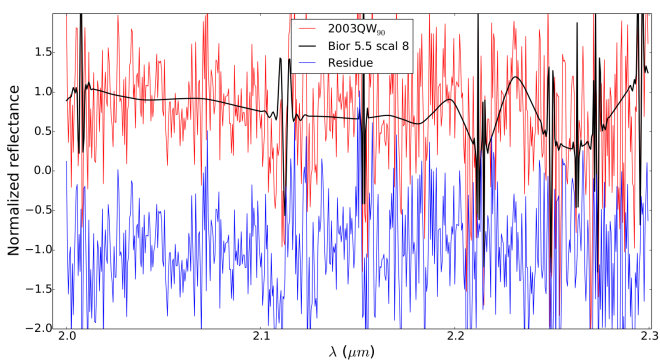}
\caption{Zoom-in of the region between 1.9 and 2.3 $\mu$m of TNO 2003 QW$_{90}$. The original data is shown in red, the filtered spectrum with the wavelet Bior 5.5 (scale 8) is shown in black, and the difference between the original and filtered spectra is shown in blue. The residue distribution was shifted downward by 1 unit for better visualization.}
\label{zoomqw90}
\end{figure}

Figure \ref{tno_p} shows how the technique works on the spectra of smaller TNOs, for instance the spectrum of 2003 QW$_{90}$. In this case the optimal scale for filtering according to the algorithm is the Bi-orthogonal 5.5 in scale 8. Analyzing  Fig. \ref{tno_p}, we see that although the amount of noise is large, the technique did not insert any obvious artifact when trying to filter the spectrum. The amount of noise decreased, but some noise still remains in the spectrum. We conclude that the wavelet simple thresholding has a limit: it filters to the point where it is no longer possible to remove noise without removing signal from the spectrum. Analyzing the bottom box, we see that the residue is larger than in the previous case, but there are still no obvious structures or patterns in it, as can be seen in Fig. \ref{zoomqw90} where we show a zoom of the final region of the spectrum of this TNO.

\subsection{Limitations}
It can be seen that, whenever the spectra have  high S/N$_o$, the absorption structures are very well respected. In the case of lower S/N$_o$ this might not strictly be the case  because the wavelets' optimal performance assume a signal that is  contaminated by white (Gaussian) noise.

Our data was obtained by photon counting, and  is thus contaminated by Poisson noise. This is not a big problem since in the limit of large numbers the Poisson distribution tends to the Gaussian  (see \citealt{bevington2003data}). More complicated are systematic noises (instrumental and calibration), which do not necessarily follow a statistical distribution and will affect any filtering technique. Examples of this can be seen in Figs. 1 and 2 with a feature close to 2.2 $\mu$m due to incomplete removal of telluric features.

\vspace{0.5cm}
In summary, through this analysis we conclude that the application of the wavelets technique worked well to filter out TNO spectra in both cases (low and high signal-to-noise ratios); therefore, we will assume it works just as well for the intermediate cases. Now, we  show the results of the comparisons between these de-noised spectra and the spectra of water and methanol ices in order to find evidence of these ices on the surfaces of these objects. 

\subsection{Comparison}

\subsubsection{Water ice}

\begin{figure}[!h]
\centering
\includegraphics[height = 5cm]{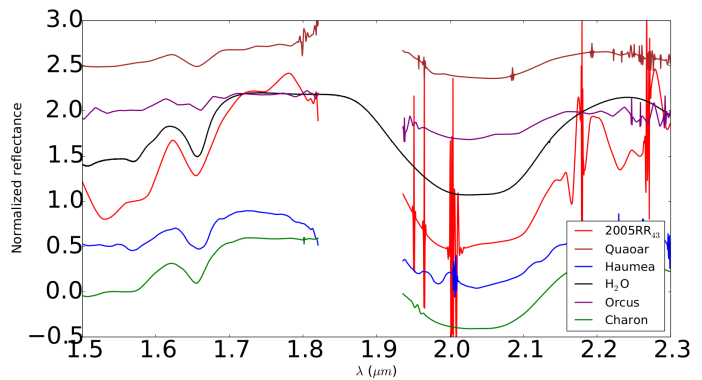}
\caption{Normalized reflectance vs. wavelength. The spectra were shifted in flux for better visualization. The region close to 1.85 $\mu$m was masked due to the absorptions of the atmosphere. In green the spectrum of Charon, in blue Haumea, in red 2005 RR$_{43}$, in purple Orcus, and in brown, Quaoar. There is a  coincidence of the band at 1.65 and 2 $\mu$m in all spectra of TNOs, which are signatures of the water ice in these objects.}
\label{agua_g}
\end{figure}

According to Table 1 the candidates in our sample that possess water ice are Quaoar, Orcus, 2003 FY$_{128}$, Haumea, Sedna, 2005 RR$_{43}$, and Charon (all with D$_{2.0}>$ 25\%). With shallower bands we have 2003 AZ$_{84}$, 2005 RM$_{43}$, 1996 TL$_{66}$, 2004 UX$_{10}$, 2003 UZ$_{413}$, Triton, Crantor, and Huya (15 $<$ D$_{2.0} <$ 25\%). Among these, those that showed possible evidence through direct comparison with the water ice spectrum  were Quaoar, Orcus, Haumea, 2005 RR$_{43}$, and Charon. The amount of noise remaining in the spectra of the other TNOs did not allow the identification of the possible water ice band. However, the drop in flux in the 2 $\mu$m region is indicative that some absorption is occurring in the region. Figure \ref{agua_g} shows spectra of objects displaying clear water ice features. 

For large TNOs we  note the similarity between the water ice spectrum and the object spectra. The band at 1.65 $\mu$m is conspicuous in all spectra. This band indicates the presence of  crystalline structure ice on the surface of TNOs, which might be related to a recent exposition of the ice that has had no time to become amorphous due to space weathering. The band at 2 $\mu$m appears in all spectra, but with different depths.

\subsubsection{Methanol ice}

\begin{figure}[!h]
\centering
\includegraphics[height = 5cm]{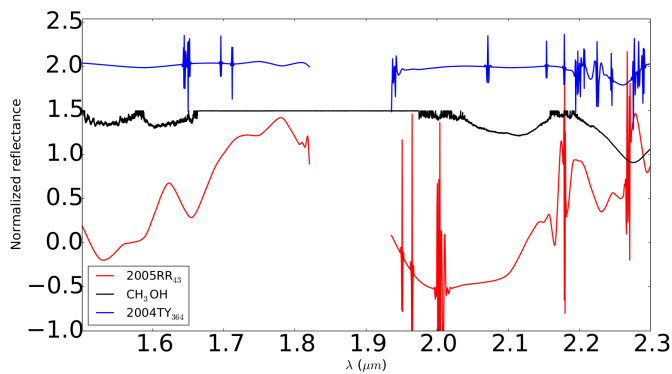}
\caption{Normalized flux vs. wavelength. The spectra were shifted in flux for better visualization. From top to bottom the spectra shown are 2004 TY$_{364}$ in blue, the model of methanol (CH$_3$OH) in black, and 2005 RR$_{43}$ in red.}
\label{metanol_g}
\end{figure}

\begin{figure}[!h]
\centering
\includegraphics[height = 5cm]{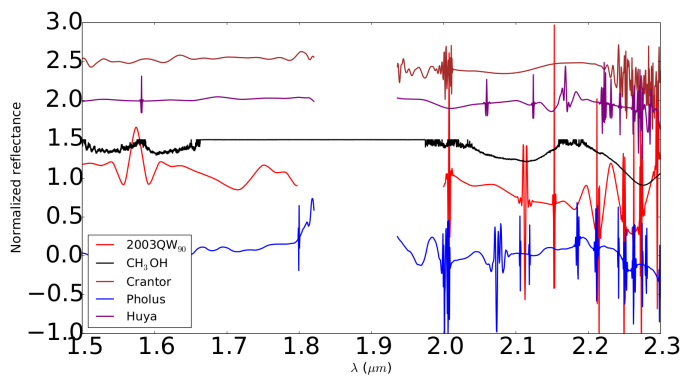}
\caption{Normalized flux vs. wavelength. The spectra were shifted in flux for better visualization. From top to bottom, the spectra shown are Crantor in brown, Huya in purple, the model of methanol (CH$_3$OH) in black, 2003 QW$_{90}$ in red, and Pholus in blue.}
\label{metanol_p}
\end{figure}

The objects in our sample with evidence of having methanol ice, according to the criterion proposed in \cite{merlin2012} and after comparison with the spectrum obtained in laboratory to methanol ice, are 2004 TY$_{364}$, 2003 QW$_{90}$, Pholus, Huya, Crantor, and 2005 RR$_{43}$. For the other objects in our sample it was not possible to find structures that could indicate the presence of this ice. In Fig. \ref{metanol_g} we show the larger objects in the sample, while in Fig. \ref{metanol_p} we show the smaller objects with possible presence of methanol in their surface composition.

Our analysis focuses on the final part of the spectrum, specifically the region beyond 2 $\mu$m. In this region the amount of noise is large and therefore is it difficult to detect methanol ice structures for any size range. However, it is possible to perceive signs as flux drop that could be associated with small and shallow absorption features. Nevertheless, care must be taken not to misinterpret some spurious features, as mentioned above.

\section{Conclusions}

In this paper, we show how the application of the wavelets allowed us to analyze the spectra of 50 trans-Neptunian objects with diameters between 50 and 2300 km. Some of the objects in our sample already have data available in the literature and were taken as validators of the technique. As the results were positive, we applied the technique for the rest of the objects in our sample.
 
We performed the filtering of the spectra of the sample in a two-step process. First we developed an algorithm that determines the best wavelet and the respective scale that optimizes the result of the filtering, and we filtered the spectra with the wavelet and scale indicated by the algorithm. After filtration, we calculated some spectral parameters to find out in which of the sample spectra there was the possibility of finding structures similar to those found in water and methanol ice spectra. We compared the filtered spectra of the candidates with the spectra calculated from optical constants measured in the laboratory for water and methanol ices. According to the similarity between the spectra, we were able to recognize the absorption bands associated with the water ice and to identify some evidence of the presence of methanol ice in some of the sample spectra. These results are summarized in Table 2.

We consider that the application of wavelets to filter the spectra was successful because it was possible to analyze TNO spectra that had previously been considered unusable due to the amount of noise they had.

\begin{table}[!h]
\centering
 \begin{tabular}{cc}
 \hline
Name & Ices \\
\hline
5145 Pholus & Methanol+\\
%19308 1996 TO$_{66}$ & Water+\\
38628 Huya & Methanol*\\
50000 Quaoar & Water+\\
%55637 2002 UX$_{25}$ & Methanol*\\
83982 Crantor &  Methanol+\\
90482 Orcus & Water+\\
%120132 2003 FY$_{128}$ & Water*\\
120348 2004 TY$_{364}$ & Methanol+\\ 
136108 Haumea & Water+ \\
145453 2005 RR$_{43}$ & Water+, Methanol*\\
307616 2003 QW$_{90}$ & Methanol*\\
%455502 2003 UZ$_{413}$ & Water*\\
Charon & Water+ \\
\hline
 \end{tabular}
\label{tab3} 
 \caption{Summary of our results: the TNOs with the respective ices. The ices marked with + have  detections in the literature, and those with * need confirmation.}
\end{table} 

We checked whether there was any  relation between dynamical grouping (populations) and the possible existence of ices. We did not find any result of relevance, due to the small number statistics.
Future developments will have to wait until new data becomes available, especially for smaller objects, when we will be able to  apply filtering by wavelets to extract the maximum information possible.

\textit{Acknowledgements.} 
This work was developed during A.C.S.F.'s M.Sc. degree at the National Observatory with financial support from CAPES. 
A.A.C. would like to acknowledge support from CNPq and FAPERJ. 
Y.J.T. acknowledges financial support from the Fundação Carlos Chagas Filho de Amparo à Pesquisa do Estado do Rio de Janeiro - FAPERJ (fellowship Nota 10, PDR-10). This research is based on observations collected at the European Organisation for Astronomical Research in the Southern Hemisphere under ESO program 178.C-0867. We thank the comments and suggestions of the referee and the editor that improved the quality of this work.

%-------------------------------------------------------------------
 \bibliographystyle{apalike}
  \bibliography{aanda}

\end{document}